\documentclass[twocolumn,aps,showpacs,preprintnumbers,superscriptaddress]{revtex4}
\usepackage{graphicx}
\usepackage{bm}
\usepackage{amsmath, amsthm, amssymb}
\usepackage{color}

\begin{document}
\title{Wave propagation, bi-directional reflectionless, and coherent perfect absorption-lasing in finite periodic PT-symmetric waveguide networks}
\author{Jeng Yi Lee}
\affiliation{Department of Opto-Electronic Engineering, National Dong Hwa University, Hualien 974301, Taiwan}

\author{Pai-Yen Chen}
\affiliation{Department of Electrical and Computer Engineering,
University of Illinois at Chicago, Chicago, Illinois 60661, USA}

\date{\today}

\begin{abstract} 
We theoretically and numerically investigate the scattering behavior of a periodic parity-time (PT)-symmetric waveguide network composed of a finite number of  unit cells.
Specifically, we put forward rigorous and formally exact expressions for wave propagation, bi-directional reflectionless, and coherent perfect absorption and lasing (CPAL) occuring in a finite periodic optical waveguide network.
Through the use of the generalized parametric space derived from observation of PT-symmetric transfer matrix, Lorentz reciprocity theorem and non-imaginary Bloch phase,
we observe that when the unit cell is operated at the PT broken phase or  exceptional point, the system can always have propagating modes, independent of the number and transmission phase of the unit cell.
On the other hand, when the unit cell is operated at the exact PT-symmetric phase, the formation of propagating waves would depend on the transmission phase of the unit cell.
More interestingly, we find that even though the unit cell is not operated at the exceptional point,  reflectionless with bi-directionality as well as unity transmittance can be achieved by choosing  appropriate number of unite cells and specific PT phases.
We also find two approaches to implement CPAL.
One is to exploit odd number of the unit cell operated at the CPAL point. 
Another way is to manipulate specific broken phase with an appropriate number of the unit cells, while making transmission phase to be null.
We believe this work may offer a theoretical underpinnings for studying  extraordinary wave phenomena of PT-symmetric photonics and may open avenues for manipulation of light.
\end{abstract}
\pacs{ }

\maketitle

\section{Introduction}
Inspired by an extension of quantum mechanics \cite{bender}, non-Hermitian systems with parity-time (PT) symmetry, had been studied in a variety of physical wave, such as photonics \cite{review2}, acoustics \cite{acoustic1,acoustic2, acoustic3}, electric circuits \cite{circuit3, circuit1,circuit2,circuit4}, elastic plate \cite{elastic1,elastic2}, coupled mechanical oscillators \cite{mechanic1,mechanic2} to name a few.
Such systems characterized by a non-Hermitian Hamiltonian, violated by unitary relation, remain invariant under a combined PT operations.
One dimensional photonics with parity-time (PT) symmetry would demand that the real part of refractive index in spatial placement has an even symmetry, while the imaginary part has an odd symmetry, corresponding to gain (amplification) and loss media (attenuation) embedded \cite{review1}.
The principal studying schemes in one-dimensional PT-symmetric photonics could be classified into longitudinal and transverse.
 In the longitudinal scheme, the propagating wave well described by paraxial equation of diffraction (Schrodinger-like equation) as well as coupled-mode formalism, would simultaneously encounter gain and loss media.
 It had been theoretically  studied and experimentally demonstrated with  coupled optical waveguides based on Fe-doped LiNbO$_{3}$ and AlGaAs \cite{pt1,pt2,pt3}. 
 In the transverse scheme, the propagating wave, described by transfer matrix or scattering matrix, would sequentially experience gain and loss media\cite{pt4,pt5,pt6,pt7}.
Although it seems that the foregoing systems are intrinsically distinct, the description of wave interactions in PT symmetry systems can be formula-equivalent to $S^{*}(\omega)=PTS(W)PT=S^{-1}(\omega)$ where $S$ is scattering matrix, P denotes inversion and $\pi$ rotation operator, T denotes antilinear complex conjugation operator, and $\omega$ is frequency \cite{pt5}.
The scattering behaviors can be  generally categorized into the PT symmetry and broken symmetry phases.
In between the symmetry and broken symmetry phases, there exists an exceptional point with scattering eigenvalues and eigenstates coalesced, corresponding to an onset of symmetry-breaking transition.

At an exceptional point, the system can display the exotic  uni-directional transparency (also known as anisotropic transmission resonance) \cite{pt7} or bi-directional transparency \cite{pt21,pt24}.
We note that the delay time, defined as the reflection coefficient with respect to operating frequency, behaves like a delta function at an exceptional point, due to an absence of reflected wave  \cite{pt32}. 
This leads to one-way (unidirectional transparency) or two-way (bi-directional transparency) light-trapping resonance in PT-symmetry systems, which  the wave would be trapped for a long time until complete absorption by lossy materials. 
Moreover, the bifurcation of scattering eigenvectors near the exceptional point reveals the potential applications to design enhanced photonic sensors \cite{pt23,pta1}.
Recent studies further  unveil that the emergence of the exceptional point is not a signature of PT-symmetric systems, while it is associated with non-Hermitian systems \cite{review2, pt22,pt23}.

Recently, we observed the possibility of symmetry reflection coefficients upon two opposite incidences, when the system is operated at the specific PT symmetry phase or exceptional point, even though PT-symmetric systems violates parity alone \cite{pt24}.
Although the occurrence of mixed PT symmetry-and broken symmetry- phases can be found in Disk PT scatterers contributed by multipolar mode resonances \cite{pt5}, the similar mixed PT phases can be  observed in the simple one-dimensional PT slabs under oblique incidences. Moreover, these properties and the corresponding  exceptional points for TE and TM polarized can be manipulated by engineering wave impedances \cite{mixedphase}. 

PT systems operated at these distinctive PT phases can exhibit exotic scattering properties, including uni-directional invisibility \cite{pt7,pt25,pt26,pt27}, subdiffraction imaging\cite{pt8,pt28}, coherent perfect absorber-laser \cite{pt5,pt6}, PT-symmetric whispering gallery mode resonators \cite{pt9,pt10},  uni-directional optical pulling force \cite{pt11,pt12,pt13}, superior sensing capability \cite{pt14,circuit1,circuit2,pt17}, single-mode microrings lasing \cite{pt18,pt19}, and Bloch oscillations \cite{pt20}.
\begin{figure*}[htp]
\centering
\includegraphics[width=0.9\textwidth]{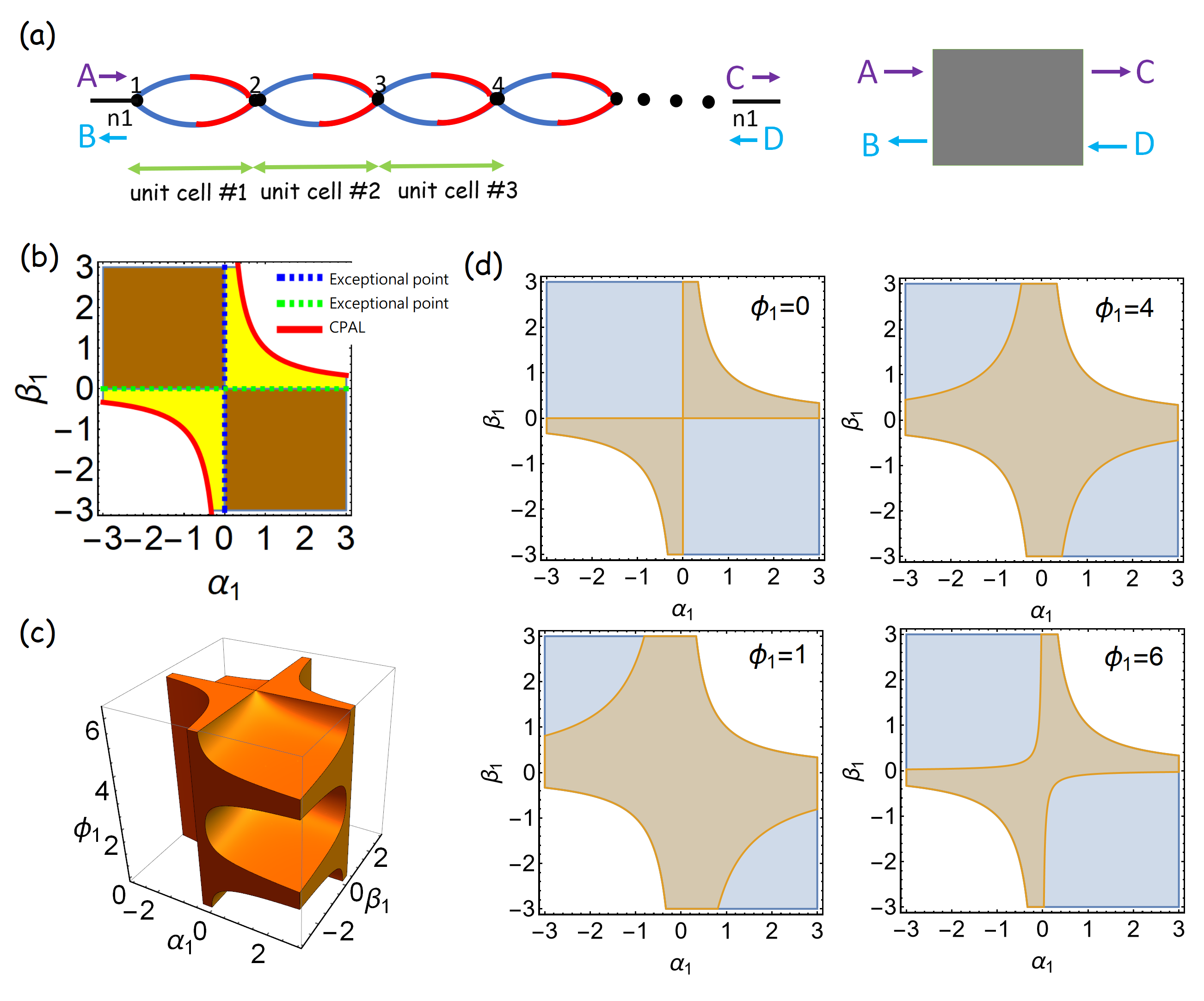}
\caption{(a) Schematic of an optical waveguide network composed of finite periodic PT-symmetric unit cells. In the unit cell, it is composed of two segments of waveguides. Each semgment is made of gain and loss waveguides connected, as marked by red and blue lines, respectively. The total arc-lengths of gain and loss waveguides are $\frac{l}{2}$ and the corresponding refractive indexes for gain and loss are $n_g$ and $n_l=n_g^{*}$, respectively. In the left and right leads, it is constructed by lossless waveguides  with refractive index $n_1$. Here $A$ and $B$ ($C$ and $D$) are complex amplitudes of scalar plane waves in the left lead (right lead) toward right and left propagation, respectively. Description of wave propagation in our waveguide network can be equivalent to two-port systems as shown in the right panel. In (b), we plot a parametric space valid for any PT-symmetric unit cells, derived from Lorentz reciprocity theorem and PT-symmetric transfer matrix. We mark symmetry phase, broken symmetry phase, exceptional point, and CPAL by brown region, yellow region, blue dashing line, green dashing line, and red solid line, respectively. The white region denotes forbidden parametrization for any PT-symmetry unit cells. Then combined with real number of the Bloch phase, i.e., Eq.(3),  in terms of parametrization $\alpha_1$, $\beta_1$, and $\phi_1$, we can construct a 3D plot with orange color in (c) to show accessible parametrization supporting propagating waves in finite periodic PT-symmetry systems. By taking transmission phases $\phi_1=0,1,4,6$, we show the region of the non-imaginary complex values of the Bloch phase $\Phi$  in the light orange region of the parametric space (d). We note that although the parametrization in the light blue region is accessible for general PT-symmetry systems, it corresponds to the complex number of Bloch phase.}
\end{figure*}
However, most of these works focused on a simple gain-loss architecture. 
We note that although there have been some efforts dedicated to  periodic PT-symmetric structures of Refs.\cite{pt7,pt31,pt26,pt29,pt30, pta2}, the relationship between PT/Bloch phases and the unit cell, especially for the finite periodic arrangements, have not yet been explored.
Moreover, in this work, we derive the criteria expressions to determine exotic scattering properties, such as coherent perfect absorption and lasing (CPAL) and bidirectional transparency, dependence of cell number.

On the flip side, thanks to the intrinsic ultrafast speed and ultralow energy consumption in nanophotonics, optical neural networks provide alternative opportunity to replace electronic architectures that have been experimentally demonstrated with silicon-based integrated circuits composed of  array of Mach–Zehnder interferometers, semiconductor laser, and photodetectors \cite{deep1} to perform matrix-vector multiplications.
Moreover, optical neural networks with embedding  gain-and-loss couplers with PT symmetry exhibit a better training performance on the digit recognition task compared to that of conventional ones \cite{deep}.
Inspired by these works and the state-of-art of III-V semiconductor photonics, we 
employ the finite periodic optical waveguide networks made of  PT-symmetry unit cells as a platform to discuss the scattering behaviors, which may have opportunity headed toward a optical chip laboratory.
By means of the parametric space derived from consideration of PT-symmetric transfer matrix, Lorentz reciprocity theorem and non-imaginary Bloch phase, we not only interpret the PT phase of the unit cell, but also discuss the formations of wave propagation, bi-directional transparency, and CPAL.
We could observe that when the unit cell is operated at the broken  phase or an exceptional point, the resulting integrated system would enable wave propagation, independent of total number and transmission phase of the unit cell.
However, it is not the case for the unit cell operated at the symmetry phase.
Moreover, we find that even through unit cells is not operated at exceptional point, by embedding a proper number of cells and operation of specific PT-phase,the system can still exhibit bi-directional transparency.
In addition, to implement CPAL, there are two approaches.
One to design the unit cell with the CPAL property, while its construction number has to be odd. 
Another approach is through choosing a proper number of unit cells as well as operation of specific broken symmetric phase with null transmission phase.
We believe this work may provide an alternative method to realize extraordinary waves by using a finite periodic structure.

\begin{figure*}[htp]
\centering
\includegraphics[width=1\textwidth]{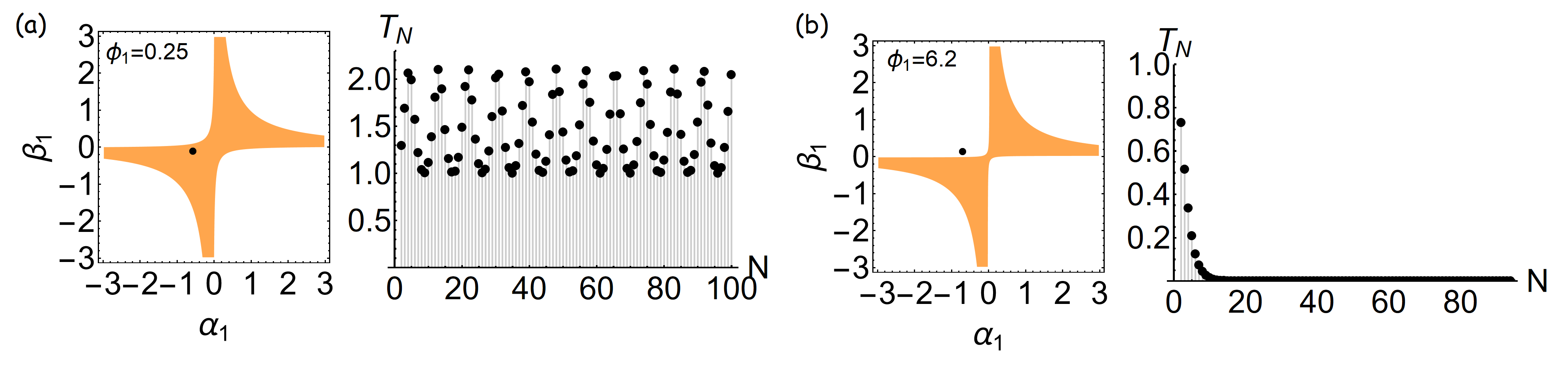}
\caption{ By means of the parametric space, we mark the non-imaginary value of the Bloch phase depicted by light blue region in the left panels of (a) and (b), with $\phi_1=0.25$ and $\phi_1=6.27$, respectively.
We design an optical waveguide system by $n_1=1.5$, $n_l=3.1+0.7i$, and $k_0=1$, but $l=2.2$ in (a) and $l=2.1$ in (b), while the corresponding PT phase 
of the unit cell is marked by a black dot.
In (a), the unit cell is operated at  broken symmetry phase, while the corresponding Bloch phase is a real number. It denotes that incident waves can propagate through any finite periodic N-cell systems. We also calculate the transmittance $T_N$ with N cells in the right panel, reflecting the propagating result.
 In (b), this unit cell is operated at symmetry phase, while the corresponding Bloch phase is a complex value. In the right panel, we can see that with increasing N-cells, $T_N$ approaches zero. This result reflects the non-propagating wave result.}
\end{figure*}

\begin{figure*}[htp]
\centering
\includegraphics[width=1\textwidth]{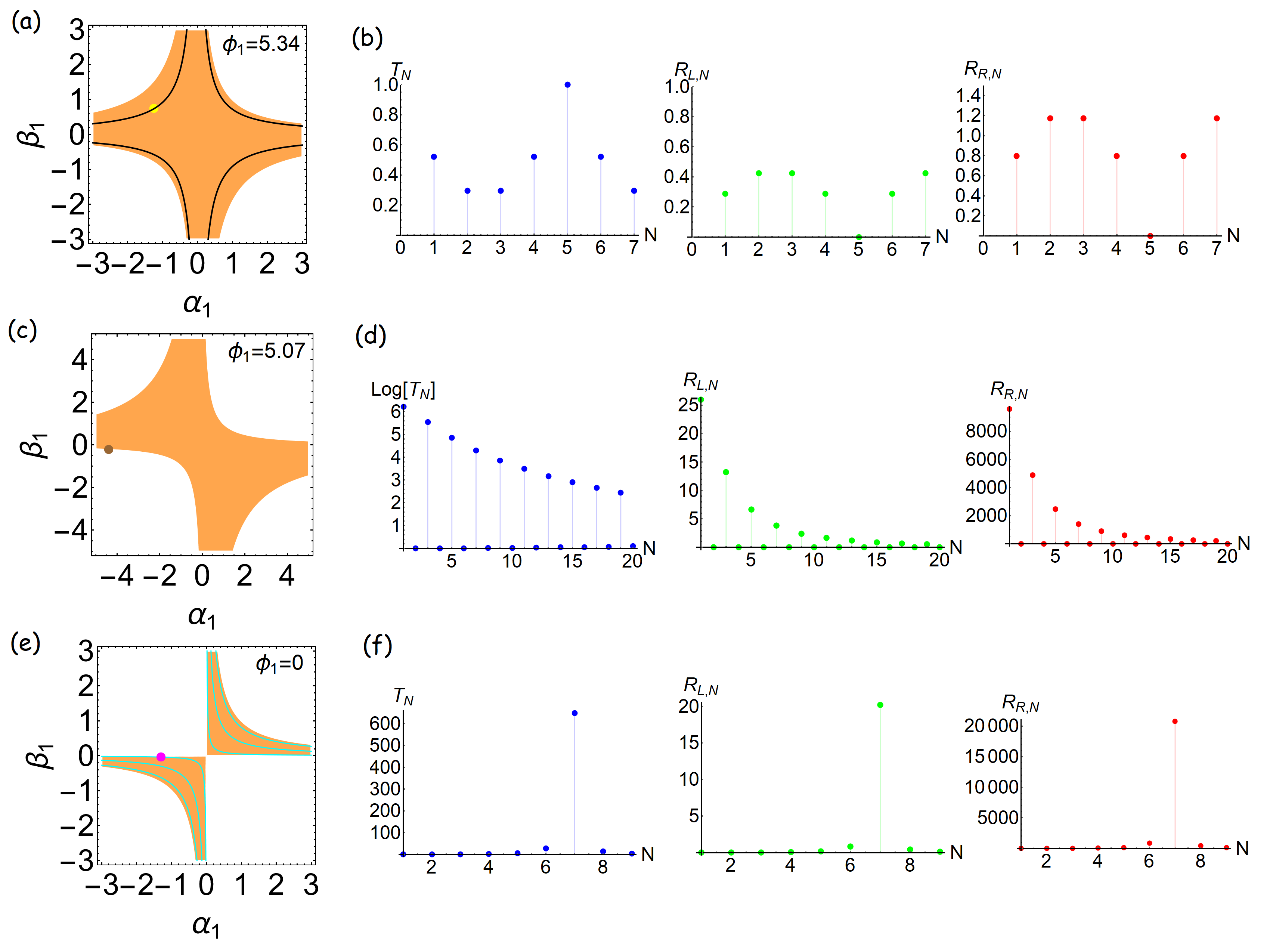}
\caption{(a) Realization of the exceptional points at finite periodic PT-symmetric systems. Obviously, this unit cell is operated at the symmetry phase and the corresponding transmission phase is $\phi_1=5.34$. (b) The corresponding transmittance and two reflectances with N. Here the parameters we use are $n_1=1.5$, $n_l=3.1+0.7i$, $k_0=1$, and  $l=1.8$, whose the PT phase is depicted by a yellow dot. The black lines are the solutions of $\sin[5\Phi]=0$ occurred at the symmetry and broken symmetry phases. We can see that two reflectances are symmetry at  $N=5$, while all reflectances become null and $T_N=1$, supporting the exceptional point with bi-directional reflectionless. In (c), we realize CPAL in finite periodic systems, while its unit cell is operated at CPAL. The transmission phase of the unit cell is $\phi_1=5.07$. When we employ this unit cell as the construction, with odd N number, $T_N$, $R_{L,N}$, and $R_{R,N}$ become huge in (d). This result supports  Eq. (6). Here the parameters are $n_1=2$, $n_l=3.1+0.5i$, $k_0=1$, and  $l=6.7$. In (e), we consider another solution to CPAL.
The unit cell is not operated at CPAL, but with specific broken symmetry phase and construction number $N=7$ to meet $\cos[7\Phi]=0$, the resultant system made of this unit cell can perform CPAL. In (f), we calculate the $T_N$, $R_{L,N}$, and $R_{R,N}$ with N. We note that this solution  is required to have the null transmission phase, i.e., $\phi_1=0$, in the unit cell. Here the parameters we use in (e) and (f) are $n_1=0.85$, $n_l=1.61+1.49i$, $k_0=1$, and  $l=0.75$. }
\end{figure*}

\section{ Theory}

\subsection{Transfer matrix of finite periodic PT-symmetric systems}
 We consider a PT-symmetric periodic structure by an optical waveguide network, as shown in Fig. 1 (a).
Its construction is formed by finite periodic unit cells in which there are two separated waveguide segments.
These segmental waveguides are operated at single-mode propagation and in an absence of wave coupling and bending loss.
The simplification of non-coupling of wave would not be loss of any generality, since the transfer/scattering matrix for the description of a PT-symmetry coupler is mathematical equivalent to our case in following discussion.

Now, in our system, each segment is made of two sub-segmental waveguides connected where one is gain and another is loss.
Here the corresponding refractive indexes for gain and loss are $n_g$ and $n_l=n_g^{*}$, respectively, and the lengths for each sub-segmental waveguides are $\frac{l}{2}$, in order to  meet PT-symmetry condition.

To have gain materials, there can conventionally achieved by electric pumping or optical pumping.
For the former, by adjusting electric carrier injection in III-V semiconductors, one can modulate the gain parameters.
For the latter, it can be nanostructured by quantum dots, quantum wells, dye molecules,  halide perovskites and graphene metasurface \cite{active,pt14}. 

We let the transfer matrix of the unit cell be $M_1$, and due to PT-symmetry embedded, it obeys $M_1^{*}=M_1^{-1}$. 
In addition, since the left and right leads of the optical waveguide network are connected to identical lossless waveguides, according to Lorentz reciprocity theorem, we have $Det[M_1^N]=1$. 
Here $N$ denotes total number of unit cells.
The refractive index for the lossless waveguide is $n_1$.
Moreover, since $Det[M_1^N]=1$ is valid for any N, we can argue that it would lead to $Det[M_1]=1$. 
Combined with $M_1^{*}=M_1^{-1}$ and $Det[M_1]=1$ for the unit cell, we can also find $(M_1^N)^{-1}=(M_1^N)^{*}$.
The detailed analysis is placed in appendix A.
This result means that the whole optical waveguide network has PT-symmetry invariant.
The transfer matrix for the unit cell can be parametrized by
\begin{equation}
\begin{split}
M_1=
\begin{bmatrix}
\sqrt{1-\alpha_1\beta_1}e^{i\phi_1} & i\alpha_1\\
i\beta_1 & \sqrt{1-\alpha_1\beta_1}e^{-i\phi_1}
\end{bmatrix}
\end{split}
\end{equation}
here $\alpha_1$ and $\beta_1$ are reals and $\phi_1$ is transmission phase bound by $[0,2\pi)$ , Refs \cite{pt24,pt13}.
The exact description of $\alpha_1$, $\beta_1$, and $\phi_1$ would depend on system configurations.
The use of parameterization is valid not only for our case, but also for a variety of PT-symmetry wave physics.
In appendix B, we derive the corresponding transfer matrix for our optical waveguide networks.

Due to Lorentz reciprocity theorem established in the unit cell, i.e., $Det[M_1]=1$, there has a constraint for $\alpha_1$ and $\beta_1$, i.e., $\alpha_1\beta_1\leq 1$  \cite{pt24,pt13}.
For an optical waveguide network composed of identical N-unit cells, with the help of Chebyshev identity, the corresponding transfer matrix reads
 \begin{widetext}
\begin{equation}
\begin{split}
M_1^{N}&=
\begin{bmatrix}
\sqrt{1-\alpha_1\beta_1}e^{i\phi_1} & i\alpha_1\\
i\beta_1 &\sqrt{1-\alpha_1\beta_1}e^{-i\phi_1}
\end{bmatrix}^{N}\\
&=
\begin{bmatrix}
\sqrt{1-\alpha_1\beta_1}e^{i\phi_1}U_{N-1}-U_{N-2}& i\alpha_1 U_{N-1}\\
i\beta_1 U_{N-1} &\sqrt{1-\alpha_1\beta_1}e^{-i\phi_1}U_{N-1}-U_{N-2}
\end{bmatrix}=\begin{bmatrix}
\frac{1}{t_N^{*}} & \frac{r_{R,N}}{t_N}\\
-\frac{r_{L,N}}{t_N} & \frac{1}{t_N}
\end{bmatrix}
\end{split}
\end{equation}
 \end{widetext}
where $t_N$ is transmission coefficient, $r_{R,N}$ is right reflection coefficient, $r_{L,N}$ is left reflection coefficient, $U_{N}=\frac{\sin[(N+1)\Phi]}{\sin\Phi}$, and $\Phi$ is the Bloch phase \cite{book1}.
The Bloch phase is related to the eigenvalues ($\lambda_1$ and $\lambda_2$) of $M_1$,
\begin{equation}
\cos\Phi=\frac{1}{2}(\lambda_1+\lambda_2)=\frac{1}{2}Tr[M_1]=\sqrt{1-\alpha_1\beta_1}\cos[\phi_1].
\end{equation}
The occurrence of $Im[\Phi]=0$ enables incident waves propagating through the systems.

\subsection{Parametric space of unit cells}

We now turn to discuss the general scattering behaviors from any two-port systems with PT-symmetry, whose the transfer matrix is defined as $M$.
The eigenvalues of the scattering matrix $S$, which is linear relation between outgoing and incoming waves, can be expressed as $S_{\pm}=\frac{e^{iArg[M_{11}]}}{\sqrt{1-Im[M_{12}]Im[M_{21}]}}(1\pm\sqrt{Im[M_{12}]Im[M_{21}]})$\cite{pt24,pt13}.
The transmission and reflection coefficients from right or left incidences are $t=\frac{e^{iArg[M_{11}]}}{\sqrt{1-Im[M_{12}]Im[M_{21}]}}$,  $r_R=\frac{Im[M_{12}]e^{i(Arg[M_{11}]+\frac{\pi}{2})}}{\sqrt{1-Im[M_{12}]Im[M_{21}]}}$, and  $r_L=\frac{Im[M_{21}]e^{i(Arg[M_{11}]-\frac{\pi}{2})}}{\sqrt{1-Im[M_{12}]Im[M_{21}]}}$, respectively. 
In the PT-symmetric phase, the eigenvalues of scattering matrix are unimodular in magnitudes and distinguishable, while the transmittance has $T<1$. Here $T=\vert t\vert^2$.
At the symmetry phase, it requires $Im[M_{12}]Im[M_{21}]<0$.
In the broken PT-symmetric phase, the scattering eigenvalues become a reciprocal pair in magnitudes, while $T>1$.
It requires $0<Im[M_{12}]Im[M_{21}]\leq 1$.  
The system with these eigenvectors would exhibit either amplification or attenuation.
On the other hand, the eigenvalues can also exist for extreme zero and infinity occurred at the broken symmetric phase, supporting a hybrid functionality of CPAL.
It corresponds to $Im[M_{12}]Im[M_{21}]=1$.
We note that the extrinsic scattering behavior can not perform CPAL simultaneously, but it can be switchable by proper incoming waves.
PT-symmetric systems with CPAL can display highly sensitivity to detect extremely small admittance perturbations over standard electromagnetic sensors \cite{pt14,circuit1,circuit2,pt17}.
Between symmetry and broken symmetry phases, there has the exceptional point with degenerate eigenvalues $S_{+}=S_{-}=e^{iArg[M_{11}]}$. 
The system can exhibit unity transmittance as well as unidirectional reflectionless ($R_R=\vert r_R\vert^2=0$ or $R_L=\vert r_L\vert^2=0$) or bidirectional reflectionless, ($R_R=R_L=0$), also so-called unidirectional or bidirectional transparency.
It can be $Im[M_{12}]=0$, $Im[M_{21}]=0$, or $Im[M_{12}]=Im[M_{21}]=0$, respectively.
As our demonstration in Refs.\cite{pt24,pt13}, only $Im[M_{12}]$ and $Im[M_{21}]$ are responsible for symmetry phase, broken symmetry phase, and an exceptional point, irrespective of transmission phase of unit cells, i.e., $Arg[M_{11}]$.
As a result, we provide a parametric space valid for any PT-symmetric unit cells with $Im[M_{12}]\Rightarrow \alpha_1$ and $Im[M_{21}]\Rightarrow \beta_1$.
Furthermore, we indicate symmetry phase, broken symmetry phase, exceptional point, and CPAL marked by brown region, yellow region, blue dashed line, green dashed line, and red solid line, respectively, as shown in Fig. 1 (b).
The parametric space provides the necessary PT phase information for  any PT-symmetry systems, irrespective of system configurations, materials, and operating frequency.
We note that the white region represents inaccessible parametrization for any PT systems.

\section{Numerical results and discussion}

\subsection{Propagating wave}
 The absence of the imaginary part of the Bloch phase, i.e.,$Im[\Phi]=0$, denotes that incident waves enable to propagate through the systems.
We should note that Eq. (3) reveals that the Bloch phase depends on  PT phase and transmission phase of the unit cell.
With Eq.(3), we provide a 3D plot to display solutions supporting propagating waves, i.e.,$Im[\Phi]=0$, marked by orange color in Fig. 1 (c).
Interestingly, not every PT phase can support propagating waves.

More specifically, we mark the non-imaginary value of the Bloch phases by using  $\phi_1=0,1,4,6$ in Fig. 1 (d).
We can observe that some regions of the symmetry phase would have complex numbers of the Bloch phase, as indicated by light-orange color.
In the broken symmetry phase and exceptional point, it requires $0<\alpha_1\beta_1\leq 1$ and $\alpha_1\beta_1=0$, respectively. 
These conditions would have non-imaginary values of the Bloch phase, because $0\leq\sqrt{1-\alpha_1\beta_1}\leq 1$.
Thus, any PT-symmetric unit cells operated at the broken symmetry phase or exceptional point would always have propagating waves in any N-cell systems, independence of transmission phase of the unit cell.
However, in the symmetry phase, it corresponds to $\alpha_1\beta_1<0$, so the factor of Eq.(3) would be subject to $\sqrt{1-\alpha_1\beta_1}>1$, leading to the occurrence of complex numbers of the Block phase, as clearly shown in the light blue color of Fig. 1 (d).
This means that an incident wave propagating through a  system composed of such unit cell would become an evanescent wave.

To demonstrate our findings, we use $n_1=1.5$, $n_l=3.1+0.7i$ and $k_0=1$ in our finite periodic waveguide network as in Figs. 2 (a) and (b).
In the Fig. 2 (a), we use $l=2.2$. 
This set of parameters has an absence of the imaginary part of the Bloch phase.
Moreover, the unit cell is operated at the broken symmetry phase.
Now, we  calculate the resultant transmittance $T_N$ with $N$, in the right panel of Fig. 2 (a).
This case supports that incident waves can propagate through the system  for any value N.
In Fig. 2 (b), we use $l=2.1$.
From the parametric space, we can see that this set of parameters corresponds to a complex number of the Bloch phase and the unit cell is operated at symmetry phase.
In the right panel of Fig. 2 (b), we calculate $T_N$ with N, supporting non-propagating wave results.

\subsection{Transmittance relation}
To figure out the transmittance relation between the unit cell and the finite periodic system,  \cite{pt31} had proposed
\begin{equation}
1-\frac{1}{T_N}=(1-\frac{1}{T_1})\frac{\sin^2[N\Phi]}{\sin^2\Phi}
\end{equation}
, obtained from an extension of generalized conservation relation.

Now, due to non-negative property of $\frac{\sin^2[N\Phi]}{\sin^2\Phi}\geq 0$, we can further obtain $T_N\leq 1$, when the unit cell operated at the symmetry phase with $T_1<1$.
This result indicates that the scattering behaviors from any finite periodic systems made of the unit cells at symmetry phase would always have symmetry phase or exceptional point, for any N.
Now, if the unit cell is operated at the exceptional point, it obeys $T_1=1$.
Again, with non-negative property of $\frac{\sin^2[N\Phi]}{\sin^2\Phi}\geq 0$, we can obtain $T_N=1$. 
Thus, any periodic systems composed of the unit cells with the exceptional points would always be at the exceptional point.
For the unit cell operated at the broken symmetry phase, we can obtain $T_N\geq 1$ due to that $T_1>1$.
As a result, any periodic systems made of the unit cells at broken symmetry phase would eventually perform the broken symmetry or exceptional point, as indicated in Fig. 2 (a).

\subsection{Exceptional point}
Any PT-symmetric systems operated at the exceptional point would accompany by unidirectional or bidirectional reflectionless, as well as unity transmittance \cite{pt24,pt13}. 
The corresponding transmission phase can be arbitrary, so PT-symmetric systems operated at the exceptional points can not be regarded as rigorous invisibility \cite{pt21}.
Any unit cells at the exceptional point can have $(M_1)_{12}=0$ or $(M_1)_{21}=0$ or $(M_1)_{12}=(M_1)_{21}=0$, corresponding to $\alpha_1=0$ or $\beta_1=0$ or $\alpha_1=\beta_1=0$.
By employing these conditions to $M_1^N$, it can lead to the resultant periodic PT-symmetric systems having the exceptional point.
Another possibility to have the exceptional point in $M_1^N$, in the absence of the unit cell operated at the exceptional point, is
\begin{equation}
\sin[N\Phi]=0, 
\end{equation}
obtained from Eq.(2), already proposed by Refs. \cite{pt29,pt30}.
We note that this condition would support the bidirectional reflectionless, because it results in $(M_1^N)_{12}=0$ and $(M_1^N)_{21}=0$ ($r_{R,N}=0$ and $r_{L,N}=0$).

Alternatively, it is desirable to see the Eq.(5) depicted in the parametric space combined with consideration of the non-imaginary Bloch phase, which can understand the relation of PT phase and the Bloch phase between the unit cells and its finite periodic systems.
In Fig. 3 (a), we use $n_1=1.5$, $n_l=3.1+0.7i$, $k_0=1$, and  $l=1.8$.
The corresponding PT phase of the unit cell is depicted by a yellow dot, belonging to the symmetry phase and real Bloch phase.
We show the solution of $\sin[N\Phi]=0$ with $N=5$ marked by the black lines, that occur at the symmetry phase and broken symmetry phase.
We can see that our system can meet $\sin[5\Phi]=0$.
We note that the finite periodic system can eventually  perform exceptional points, even the unit cell is not at the exceptional point.
In Fig. 3 (b), we calculate the corresponding transmittance $T_N$, left reflectance $R_{L,N}=\vert r_{L,N}\vert^2$, and right reflectance $R_{R,N}=\vert r_{R,N}\vert^2$ with N.
We note that two reflectances for the unit cell are asymmetry, except when $N=5$, we find $T=1$ and $R_{L,N}=R_{R,N}=0$, supporting the symmetric bidirectional transparency.

Thus, to achieve bidirectional transparency, the unit cell can have any values of transmission phase while it needs specific symmetry or broken PT phases and specific N unit cells.

\subsection{Coherent perfect absorber-laser (CPAL)}
PT-symmetric systems operated at CPAL would have zero and infinity eigenvalues of the scattering matrix, corresponding to coherent perfect absorber and laser, respectively \cite{pt5,pt6}. 
To have CPAL occurred at finite periodic PT-symmetric systems, it requires $(M_{1}^N)_{11}=(M_{1}^N)_{22}=0$ and $Im[(M_{1}^N)_{12}]Im[(M_{1}^N)_{21}]=1$.
Accordingly, there are two solutions: (a) the unit cell exhibits CPAL and the corresponding construction number is odd, 
\begin{equation}
\begin{cases}
\begin{split}
\alpha_1\beta_1&=1\\
N&=1,3,5,..
\end{split}
\end{cases}
\end{equation}

 (b) The unit cell is not at CPAL, while its transmission phase is null and the Bloch phase should meet $\cos[N\Phi]=0$,
 \begin{equation}
 \begin{cases}
 \begin{split}
\phi_1&=0\\
 \cos[N\Phi]&=0.
 \end{split}
 \end{cases}
 \end{equation}
The detailed derivation is placed at appendix C.

To verify the case (a), we prepare another set of parameters for our waveguide network by $n_1=2$, $n_l=3.1+0.5i$, $k_0=1$, and  $l=6.7$ in Figs. 3 (c)-(d).
The corresponding transmission phase for the unit cell is $\phi_1=5.07$.
We analyze the PT phase and the Bloch phase for the unit cell in Fig. 3 (c). 
We find that the unit cell is almost operated at CPAL.
In Fig. 3 (d), we provide the corresponding transmittance and two reflectances with respect to N.
As expectation, when $N$ is odd, the values of $T_N$, $R_{R,N}$, and $R_{L,N}$ become huge.

To verify the case (b), the parameters $n_1=0.85$, $n_l=1.61+1.49i$, $k_0=1$, and  $l=0.75$ are used in Figs. 3 (e)-(f).
The corresponding transmission phase of the unit cell is null, $\phi_1=0$.
We analyze the PT phase of the unit cell and the Bloch phase in Fig. 3 (e), marked by a magenta dot and light blue region, respectively.
This unit cell is not operated at CPAL.
Now, we consider $\cos[N\Phi]=0$ with $N=7$ as illustrated in the bright blue lines, corresponding to CPAL in finite periodic systems.
These lines should locate at the broken symmetry phase, as expectation.
We also observe that our system can satisfy $\cos[N\Phi]=0$ by $N=7$.
In the Fig. 3 (f), we calculate the $T_N$, $R_{R,N}$, and $R_{L,n}$ with respect to N.
When $N=7$, the transmittance and two reflectances become sudden huge.
This result supports the CPAL occurred at this system.
We should stress that as discussed in the subsection of transmittance relation, the corresponding unit cells  must be operated at specific broken symmetry phase.

We should emphasize that the use of the parametric space here is not only directly from the consideration of PT-symmetric transfer matrix and Lorentz reciprocity theorem, but also with non-imaginary  Bloch phase embedded.
It could therefore provide clear information about PT phases of the unit cell, the Bloch phase, and the resultant scattering behavior of finite periodic systems.
Least but not last, an inverse method to manipulate PT-symmetric functionality in subwavelength dimension is proposed in \cite{pt24}, that it could have practical applications in design of finite periodic systems and PT-symmetric band-gap structures.
With the mathematical equivalence of wave structures, the concept of our method can be applied in other PT wave fields.

\section{Conclusion}
Considering the Lorentz reciprocity theorem,  PT-symmetry condition, and non-imaginary Bloch phase, we have derived and exploited  the generalized parametric space to analyze the scattering behavior of the PT-symmetric unit cell and a network constituted by finite number of cells. 
When the unit cell is operated at the broken symmetry phase and exceptional point, the propagating mode can exist in this periodic structures, regardless the number of cells.
However, in the exact PT-symmetry phase, the occurrence of the non-imaginary value of the Bloch phase would depend on the transmission phase of unit cell.
We also investigate the formations of bi-directional transparency and CPAL in finite periodic PT-symmetric systems.
Even the unit cell is not operated in the exceptional point, by carefully manipulating specific PT phase and choosing suitable value of N cells, the finite periodic system can exhibit bi-directional transparency.
There are two approaches to realize the self-dual CPAL device.
One is to pin odd  unit cells to the CPAL point as well as odd number of the cells, while the other is to utilize unit cells having specific broken symmetry phase.
Interestingly, for the latter case, the transmission phase has to be null.
We have illuminated the proposed concept with several design.
We believe this work can provide an  universally applicable guidelines for design and implementation of finite periodic PT-symmetric networks. 

\section*{Acknowledgements}

This work was supported by Ministry of Science and Technology, Taiwan (MOST) ($107$-
$2112$-M-$259$-$007$-MY3, $110$-$2112$-M-$259$-$005$- and$111$-$2112$-M-$259$ -$011$ -).


\begin{figure*}
\centering
\includegraphics[width=0.3\textwidth]{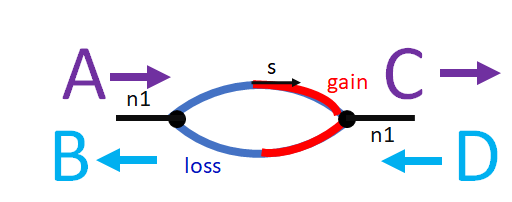}
\caption{Schematic of a unit cell in our optical waveguide network. In the left (right) lossless wveguide, $A$  and $B$ ($C$ and $D$) denote complex amplitudes of scalar plane waves toward right and left propagation. Here $s$ is arclength coordinate along the waveguides.}
\end{figure*}

 \section{APPENDIX A}
 For a unit cell with PT-symmetry, we have $M^{-1}=M^{*}$. 
 On the other hand, once the integrated system has same opposite leads, we can have $Det[M^{N}]=1$ from Lorentz reciprocity theorem.
 
In the following analysis, we would infer $Det[M]=1$ and $(M^N)^{-1}=(M^N)^{*}$, based on two  given conditions $M^{-1}=M^{*}$ and $Det[M^{N}]=1$.

Because $Det[M^{N}]=1$ is valid for any N, we argue
\begin{equation}
\begin{split}
Det[M^{N}]=Det[M]Det[M]..Det[M]&=1\\
\therefore Det[M]&=1.
\end{split}
\end{equation}

Consider $AAA^{-1}A^{-1}=I=(AA)(A^{-1}A^{-1})=A^2B$ where $A$ is invertible matrix and $I$ is unit matrix.
From linear algebra, $B=(A^{2})^{-1}$, we can find $B=(A^{-1})^2=(A^{2})^{-1}$.
Thus, $M^{-1}M^{-1}...M^{-1}=(M^{-1})^N=(M^N)^{-1}$.
Using PT symmetry, i.e.,$M^{-1}=M^{*}$, we have
\begin{equation}
\begin{split}
(M^{-1})^N&=(M^N)^{-1}\\
\therefore (M^{*})^N&=(M^N)^{*}=(M^{N})^{-1}.
\end{split}
\end{equation}
Thus we obtain $(M^{N})^{*}=(M^N)^{-1}$.

 \section{APPENDIX B}
 As shown in Fig. 4, we consider an optical waveguide network where is constructed by a finite number of unit cells. 
In each unit, there are two segments of waveguides in absences of wave coupling and bending losses
Each segment of the waveguide has two subsegments of gain and loss waveguides connected, marked by blue and red lines.
We discuss scalar wave equation and time evolution of propagating waves is $e^{-i\omega t}$.
At each connecting interface, the wave function and its derivative are continuous.
The latter condition is related to energy flux conservation, \cite{pt29,pt30}.

The corresponding wave function for the gain waveguide is 
$\psi_g(s)=ce^{ik_0n_gs}+de^{-ik_0n_g s}, 0\leq s\leq \frac{l}{2}$. 
Here $s$ is arc-length, $n_g$ is gain index of refraction, $k_0$ is vacuum wave number, $c$ and $d$ denote complex  amplitudes for right and left propagating waves. 
The total arc-lengths for the gain and loss waveguides in the unit cell are $\frac{l}{2}$. 

The corresponding wave function for loss waveguide is
$\psi_l(s)=ae^{ik_0n_ls}+be^{-ik_0n_l s}, -\frac{l}{2}\leq s\leq 0$ here $n_l$ is loss index of refraction.

At the interface, the corresponding boundary conditions are the continuity of the wave functions and its derivative, i.e., $\psi_g(0^{+})=\psi_l(0^{-})$ and $\frac{d \psi_g}{ds}(0^{+})=\frac{d \psi_l}{ds}(0^{-})$,
\begin{equation}
\begin{split}
c+d&=a+b\\
ik_0 n_g(c-d)&=ik_0 n_l(a-b).
\end{split}
\end{equation}
As a result, we can construct the transfer matrix for the  subsegmental waveguides at the interface,
\begin{equation}
\begin{split}
\begin{bmatrix}
c\\
d
\end{bmatrix}=
\begin{bmatrix}
\frac{1}{2}(1+\frac{n_l}{n_g}) &\frac{1}{2}(1-\frac{n_l}{n_g})\\
\frac{1}{2}(1-\frac{n_l}{n_g}) & \frac{1}{2}(1+\frac{n_l}{n_g})
\end{bmatrix}
\begin{bmatrix}
a\\
b
\end{bmatrix}.
\end{split}
\end{equation} 
We note that due to symmetrical geometry for upper and lower segmental waveguides as well as assumption of non-coupling wave, the wave functions for two segments in the unit cell are identical.

Now, we turn to derive the corresponding transfer matrix in the nexus, i.e., node, marked by  black dots in Fig. 4.
For node number 1, it is formed by a lossless waveguide and two lossy waveguides.
 The index of refraction for lossless waveguide is $n_1$.
The boundary conditions at the node are the continuity of wave function and its derivative, so we have
\begin{equation}
\begin{split}
\begin{bmatrix}
a\\
b
\end{bmatrix}=
\begin{bmatrix}
\frac{1}{4}(1+\frac{n_1}{n_l}) &\frac{1}{4}(1-\frac{n_1}{n_l})\\
\frac{1}{4}(1-\frac{n_1}{n_l}) & \frac{1}{4}(1+\frac{n_1}{n_l})
\end{bmatrix}
\begin{bmatrix}
A\\
B
\end{bmatrix}.
\end{split}
\end{equation}

We then derive the transfer matrix for a unit cell in the same leads, i.e., between two neighboring nodes,
 \begin{widetext}
\begin{equation}
\begin{split}
M_1=
\begin{bmatrix}
(1+\frac{n_g}{n_1}) &(1-\frac{n_g}{n_1})\\
(1-\frac{n_g}{n_1}) & (1+\frac{n_g}{n_1})
\end{bmatrix}
\begin{bmatrix}
e^{ik_0 n_g \frac{l}{2}} & 0\\
0 & e^{-ik_0 n_g \frac{l}{2}}\\
\end{bmatrix}
\begin{bmatrix}
\frac{1}{2}(1+\frac{n_l}{n_g}) &\frac{1}{2}(1-\frac{n_l}{n_g})\\
\frac{1}{2}(1-\frac{n_l}{n_g}) & \frac{1}{2}(1+\frac{n_l}{n_g})
\end{bmatrix}
\begin{bmatrix}
e^{ik_0 n_l \frac{l}{2}} & 0\\
0 & e^{-ik_0 n_l \frac{l}{2}}\\
\end{bmatrix}
\begin{bmatrix}
\frac{1}{4}(1+\frac{n_1}{n_l}) &\frac{1}{4}(1-\frac{n_1}{n_l})\\
\frac{1}{4}(1-\frac{n_1}{n_l}) & \frac{1}{4}(1+\frac{n_1}{n_l})
\end{bmatrix}
\end{split}
\end{equation}
\end{widetext}

It is straightforward to prove that $Det[ M_1]=1$.

\section{APPENDIX C}
To perform CPAL in periodic PT-symmetric systems, as our above discussion in the theory section, the necessary conditions are $(M_{1}^N)_{11}=(M_{1}^N)_{22}=0$ and $Im[(M_{1}^N)_{12}]Im[(M_{1}^N)_{21}]=1$. 
There are two solutions to meet these conditions.

(a)As the unit cell is operated at CPAL, i.e., $\alpha_1\beta_1=1$, the remaining conditions to have CPAL in periodic PT-symmetric systems are $U_{N-2}=0$ and $(U_{N-1})^2=1$, therefore,
\begin{equation}
\begin{split}
\sin[(N-1)\Phi]&=0\\
\end{split}
\end{equation}
and 
\begin{equation}
\sin[ N\Phi]=\pm 1
\end{equation}
, respectively.
To simultaneously satisfy above equations, we have $\Phi=\frac{\pi}{2},\frac{3\pi}{2}$ and $N$ is odd, $N=1,3,5,7,..$.

(b) If the unit cell is not operated at CPAL, i.e., $\alpha_1\beta_1\neq 1$, we need to require
\begin{equation}\label{eq1}
\begin{split}
Im[(M_{1}^N)_{12}]Im[(M_{1}^N)_{21}]&=1\\
\rightarrow \alpha_1\beta_1\sin^2[N\Phi]=\sin^2\Phi
\end{split}
\end{equation}

Now we turn to consider $(M_{1}^N)_{11}=0$ and $(M_{1}^N)_{22}=0$, so that it requires
\begin{equation}
\begin{split}
\sqrt{1-\alpha_1\beta_1}e^{i\phi_1}U_{N-1}&=U_{N-2}\\
\sqrt{1-\alpha_1\beta_1}e^{-i\phi_1}U_{N-1}&=U_{N-2}.
\end{split}
\end{equation}
Thus to satisfy above equations simultaneously, we need $\phi_1=0$.
Thus, the transmission phase of the unit cell is null.
Based on $\phi_1=0$, the corresponding Bloch phase would become
\begin{equation}\label{eq2}
\cos\Phi=\sqrt{1-\alpha_1\beta_1}.
\end{equation}

Then, the last remaining condition for $(M_{1}^N)_{11}=0$ and $(M_{1}^N)_{22}=0$ is
\begin{equation}
\begin{split}
&\sqrt{1-\alpha_1\beta_1}U_{N-1}-U_{N-2}\\
&=\frac{\cos\Phi\sin[N\Phi]-\sin[N\Phi]\cos\Phi+\cos[N\Phi]\sin\Phi}{\sin\Phi}\\
&=\cos[N\Phi]\Rightarrow 0.
\end{split}
\end{equation}
So we have $\cos[N\Phi]=0$.
We note that $\cos[N\Phi]=0$ is consistent with Eqs.(\ref{eq1}) an (\ref{eq2}).

In short, if the unit cell is not at CPAL, to achieve CPAL for a finite periodic PT-symmetric system, the corresponding conditions are $\phi_1=0$ and $\cos[N\Phi]=0$.


\begin{thebibliography}{99}

\bibitem{bender}
C. M. Bender and S. Boettcher, 
``Real Spectra in Non-Hermitian Hamiltonians Having PT Symmetry,''
 Phys. Rev. Lett. \textbf{80}, 5243 (1998).

\bibitem{review2}
Özdemir, K., S. Rotter, F. Nori, and L. Yang,
``Parity–time symmetry and exceptional points in photonics,''
Nat. Mater. \textbf{18}, 783–798 (2019).




\bibitem{acoustic1}
Fleury, R., D. Sounas, and A. Al\'u,
``An invisible acoustic sensor based on parity-time symmetry,''
” Nat. Commun. \textbf{6}, 5905 (2015).

\bibitem{acoustic2}
X. F. Zhu, H. Ramezani, C. Z. Shi, J. Zhu, and X. Zhang,
``PT-Symmetric Acoustics,'' Phys. Rev. X \textbf{4}, 031042 (2014).

\bibitem{acoustic3}
Y. Aur\'egan and V. Pagneux,
``PT-Symmetric Scattering in Flow Duct Acoustics,''
Phys. Rev. Lett. \textbf{118}, 174301 (2017).


\bibitem{circuit3}
J. Schindler, A. Li, M. C. Zheng, F. M. Ellis, and T. Kottos,
``Experimental study of active LRC circuits with PT symmetries,''
Phys. Rev. A textbf{84}, 040101(R) (2011).

\bibitem{circuit1}
M. Sakhdari, M. Hajizadegan, Q. Zhong, D. N.
Christodoulides, R. El-Ganainy, and P. Y. Chen,
``Experimental Observation of PT Symmetry Breaking near Divergent Exceptional Points,''
Phys. Rev. Lett. \textbf{123}, 193901 (2019).


\bibitem{circuit2}
P.-Y. Chen, M. Sakhdari, M. Hajizadegan, Q. Cui,
M. M.-C. Cheng, R. El-Ganainy, and A. Al\'u,
``Generalized parity–time symmetry condition for enhanced sensor telemetry,''
Nat. Electron. \textbf{1}, 297 (2018).


\bibitem{circuit4}
J. Schindler, Z. Lin, J. M. Lee, H. Ramezani, F. M. Ellis, and T.
Kottos,
``PT -symmetric electronics,''
, J. Phys. A: Math. Theor. \textbf{45},
444029 (2012).

\bibitem{elastic1}
M. Farhat, P.-Y. Chen, S. Guenneau, and Y. Wu,
``Self-dual singularity through lasing and antilasing in thin elastic plates,''
Phys. Rev. B \textbf{103}, 134101 (2021).

\bibitem{elastic2}
 Z. Hou, H. Ni, and B. Assouar,
``PT -Symmetry for Elastic Negative Refraction,''
 Phys. Rev. Appl. \textbf{10}, 44071 (2018).

\bibitem{mechanic1}
C. M. Bender, M. Gianfreda, and S. P. Klevansky,
``Systems of coupled PT -symmetric oscillators,''
Phys. Rev. A \textbf{90}, 022114 (2014).

\bibitem{mechanic2}
X. W. Xu, Y. X. Liu, C. P. Sun, and Y. Li,
``Mechanical PT symmetry in coupled optomechanical systems,''
 Phys. Rev. A \textbf{92}, 013852 (2015).

\bibitem{review1}
R. El-Ganainy, K. G. Makris, M. Khajavikhan, Z. H.
Musslimani, S. Rotter, and D. N. Christodoulides, 
``Non-Hermitian physics and PT symmetry,''
Nat. Phys. \textbf{14}, 11 (2018).

\bibitem{pt1}
C. E. Ruter, K. G. Makris, R. El-Ganainy, D. N.
Christodoulides, M. Segev, and D. Kip,
``Observation of parity–time symmetry in optics,''
Nat. Phys. \textbf{6}, 192 (2010).

\bibitem{pt2}
A. Guo, G. J. Salamo, D. Duchesne, R. Morandotti, M. Volatier-Ravat, V. Aimez, G. A. Siviloglou, and D. N. Christodoulides,
``Observation of PT-Symmetry Breaking in Complex Optical Potentials,''
Phys. Rev. Lett. \textbf{103}, 093902 (2009).

\bibitem{pt3}
R. El-Ganainy, K. G. Makris, D. N. Christodoulides, and
Z. H. Musslimani,
``Theory of coupled optical PT-symmetric structures,''
 Opt. Lett. \textbf{32}, 2632 (2007).

\bibitem{pt4}
L. Feng, Y.-L. Xu, W. F. Fegadolli, M.-H. Lu, J. E. B.
Oliveira, V. R. Almeida, Y.-F. Chen, and A. Scherer,
``Experimental demonstration of a unidirectional reflectionless parity-time metamaterial at optical frequencies,''
Nat. Mater. \textbf{12}, 108 (2013).
 
 
 \bibitem{pt5}
 Y. D. Chong, Li Ge, and A. Douglas Stone,
 ``PT-Symmetry Breaking and Laser-Absorber Modes in Optical Scattering Systems,''
 Phys. Rev. Lett. \textbf{106}, 093902 (2011).

 \bibitem{pt6}
 S. Longhi,
 ``PT-symmetric laser absorber,''
 Phys. Rev. A \textbf{82}, 031801(R) (2010).
 
  \bibitem{pt7}
  Z. Lin, H. Ramezani, T. Eichelkraut, T. Kottos, H. Cao,
and D. N. Christodoulides,
 ``Unidirectional Invisibility Induced by PT-Symmetric Periodic Structures,''
 Phys. Rev. Lett. \textbf{106}, 213901 (2011).
 


  \bibitem{pt21}
 L. Ge, Y. D. Chong, and A. D. Stone, ``Conservation Relations and Anisotropic Transmission Resonances in One Dimensional PT -Symmetric Photonic Heterostructures,''
Phys. Rev. A 85, 023802 (2012).

 
 \bibitem{pt24}
 J. Y. Lee and P.-Y. Chen,
 ``Generalized parametric space, parity symmetry of reflection, and systematic design approach
for parity-time-symmetric photonic systems,''
Phys. Rev. A \textbf{104}, 033510 (2021).


\bibitem{pt32}
H. Ramezani, H.-K. Li, Y. Wang, and X. Zhang, 
``Unidirectional Spectral Singularities,''
Phys. Rev. Lett. \textbf{113}, 263905 (2014). 




\bibitem{pt23}
M.-A. Miri and A. Al\'u,
``Exceptional points in optics and photonics,'' Science \textbf{363}, eaar7709 (2019).



\bibitem{pta1}
Pai-Yen Chen, Maryam Sakhdari, Mehdi Hajizadegan, Qingsong Cui, Mark Ming-Cheng Cheng, Ramy El-Ganainy, and Andrea Alù,
``Generalized parity–time symmetry condition for enhanced sensor telemetry,'' Nature Electronics \textbf{1}, 297-304 (2018).
 
\bibitem{pt22}
L. Ge and L. Feng,
``Optical-reciprocity-induced symmetry in photonic heterostructures and its manifestation
in scattering PT -symmetry breaking,''  Phys. Rev. A \textbf{94}, 043836 (2016).




\bibitem{mixedphase}
S. Droulias, I. Katsantonis, M. Kafesaki, C. M. Soukoulis, and E. N. Economou,
``Accessible phases via wave impedance engineering with PT-symmetric metamaterials,'' Phys. Rev. B \textbf{100}, 205133 (2019).
 
\bibitem{pt25}
Ali Mostafazadeh,
``Invisibility and PT symmetry,'' Phys. Rev. A \textbf{87}, 012103 (2013).

\bibitem{pt26}
 L. Yuan and Y. Y. Lu,
``Unidirectional reflectionless transmission for two-dimensional PT -symmetric periodic structures,''
Phys. Rev. A \textbf{100}, 053805 (2019).

\bibitem{pt27}
 S. Savoia, G. Castaldi, V. Galdi, A. Al\'u, and N. Engeta,
``Tunneling of obliquely incident waves through PT -symmetric epsilon-near-zero bilayers,''
Phys. Rev. B \textbf{89}, 085105 (2014).


\bibitem{pt8} 
R. Fleury, D. L. Sounas, and A. Al\'u,
``Negative Refraction and Planar Focusing Based on Parity-Time Symmetric Metasurfaces,''
 Phys. Rev. Lett. \textbf{113}, 023903 (2014).
 
 
 \bibitem{pt28}
 F. Monticone, C. A. Valagiannopoulos, and A. Al\'u
 ``Parity-Time Symmetric Nonlocal Metasurfaces: All-Angle Negative Refraction and Volumetric Imaging,'' Phys. Rev. X \textbf{6}, 041018 (2016).
 
 
 \bibitem{pt9}
 B. Peng, Ş. K. Özdemir, F. Lei, F. Monifi, M. Gianfreda,
G. L. Long, S. Fan, F. Nori, C. M. Bender, and L. Yang,
 ``Parity-time-symmetric whispering-gallery microcavities,'' Nat. Phys. \textbf{10}, 394 (2014).
 
 \bibitem{pt10}
 L. Chang, X. Jiang, S. Hua, C. Yang, J. Wen, L. Jiang, G. Li,
G. Wang, and M. Xiao,
 ``Parity-time symmetry and variable optical isolation in
active-passive-coupled microresonators,''  Nature
Photon. \textbf{8}, 524 (2014).
 
 \bibitem{pt11}
 R. Alaee, J. Christensen, and M. Kadic,
 ``Optical Pulling and Pushing Forces in Bilayer PT -Symmetric Structures,''
 Phys. Rev. Applied \textbf{9}, 014007  (2018).
 
 \bibitem{pt12}
 R. Alaee, B. Gurlek, J. Christensen, and M. Kadic,
 ``Optical force rectifiers based on PT -symmetric metasurfaces,''
 Phys. Rev. B \textbf{97}, 195420 (2018).
 
 \bibitem{pt13}
 J. Y. Lee and P.-Y. Chen,
 ``Optical forces and directionality in one-dimensional PT -symmetric photonics,''
 Phys. Rev. B \textbf{104}, 245426 (2021).
 
 
 \bibitem{pt14}
 M. Farhat, M. Yang, Z. Ye, and P.-Y. Chen,
 ``PT-Symmetric Absorber-Laser Enables Electromagnetic Sensors with
Unprecedented Sensitivity,'' ACS Photonics \textbf{7}, 2080 (2020).





\bibitem{pt17}
P. Y. Chen and J. Jung,
``PT Symmetry and Singularity- Enhanced Sensing based on Photoexcited Graphene Metasurfaces,''
Phys. Rev. Appl. \textbf{5}, 064018 (2016).


\bibitem{pt18}
H. Hodaei, M.-A. Miri, M. Heinrich, D. N. Christodoulides,
and M. Khajavikhan,
``Parity-time-symmetric microring lasers,'' Science \textbf{346}, 975 (2014). 

\bibitem{pt19}
 L. Feng, Z. J. Wong, R.-M. Ma, Y. Wang, and X. Zhang,
``Single-mode laser by parity-time symmetry breaking,''
Science \textbf{346}, 972 (2014).

\bibitem{pt20}
 S. Longhi, 
``Bloch Oscillations in Complex Crystals with  PT Symmetry,''
Phys. Rev. Lett. \textbf{103}, 123601 (2009).
 
  \bibitem{pt31}
 V. Achilleos, Y. Au\'regan, and V. Pagneux,
``Scattering by Finite Periodic PT -Symmetric Structures,''
Phys. Rev. Lett. \textbf{119}, 243904 (2017).
 
 \bibitem{pt29}
 H. Wu, X. Yang, D. Deng, and H. Liu,
 ``Reflectionless phenomenon in PT -symmetric periodic structures of one-dimensional
two-material optical waveguide networks,''
Phys. Rev. A \textbf{100}, 033832 (2019).

\bibitem{pt30}
 H. Wu, X. Yang, Y. Tang, X. Tang, D. Deng, H. Liu, and Z. Wei,
``The Scattering Problem in PT-Symmetric Periodic Structures of 1D Two-Material Waveguide Networks,''
Ann. Phys. 1900120 (2019).

\bibitem{pta2}
J. Zheng, X. Yang, D. Deng, and H. Liu, 
``Singular properties generated by finite periodic PT-symmetric optical waveguide network,''
Opt. Express \textbf{27}, 1538 (2019).

\bibitem{deep1}
Y. Shen, N. C. Harris, S. Skirlo, M. Prabhu, T. Baehr-Jones,
M. Hochberg, X. Sun, S. Zhao, H. Larochelle, D. Englund
et al.,
``Deep learning with coherent nanophotonic circuits,'' Nature photonics  \textbf{11}, 441–446 (2017)

\bibitem{deep}
H. Deng and M. Khajavikhan,
``Parity–time symmetric optical neural networ,''
Optica, \textbf{8}, 1328–1333, 2021.



\bibitem{active}
 A. Krasnok and A. Al\'u, ``Active nanophotonics,'' Proc. IEEE \textbf{108}, 628 (2020)

\bibitem{book1}
P. Markos and C. M. Soukoulis, \textit{Wave Propagation: From
Electrons to Photonic Crystals and Left-Handed Materials}
(Princeton University Press, Princeton, 2008).



\end{thebibliography}
\end{document}